\documentclass[twocolumn, letter]{jpsj3_Pre}
\usepackage{txfonts}

\title{Phase Transition to Insulating State from Quantum Hall State by Current-Induced Nuclear Spin Polarization}

\author{S. Tsuda$^1$\thanks{E-mail: shibun@scphys.kyoto-u.ac.jp}, Minh-Hai Nguyen$^1$\thanks{Present address: Department of Physics, Cornell University, Ithaca, New York 14853},
D. Terasawa$^2$, A. Fukuda$^2$, 
and A. Sawada$^3$}
\inst{$^1$Department of Physics, Graduate School of Science, Kyoto University, Kyoto 606-8502, Japan \\
$^2$Department of Physics, Hyogo College of Medicine, Nishinomiya 663-8501, Japan \\
$^3$Research Center for Low Temperature and Materials Sciences, Kyoto
University, Kyoto 606-8501, Japan\\
 
{\rm (Dated: \today)}} %\\

\abst{We investigate the resistance enhancement state (RES) where the magnetoresistance of the $\nu = 2/3$ fractional quantum Hall state (FQHS) is increased with dynamic nuclear spin polarization (DNP) induced by a large electric current.
After inducing DNP, we measure the temperature dependence of the magnetoresistance 
by a small current over a short period of time.
We find that the FQHS makes a phase transition to an insulating state.
By measuring the Hall resistance in the insulating state, we find that the RES exhibits a quantized Hall resistance.
We discuss the RES in association with the Anderson localization.}

\kword{semiconductor, quantum Hall state, fractional quantum Hall state, nuclear spin polarization, domain,
Wigner crystal, insulator, Anderson localization}

\begin{document}
\maketitle

When subjected to a strong perpendicular magnetic field at low temperatures, a two-dimensional electron system (2DES) displays novel collective states arising from the dominance of the Coulomb interaction energy over the kinetic energy and the degree of freedom of the spins. Under a high magnetic field, when the electrons occupy the lowest Landau level, the 2DES exhibits fractional quantum Hall states (FQHSs). The FQHSs show not only a variety of phenomena originating from Coulomb interaction but also spin-related phenomena due to the spin degree of freedom  \cite{Iw1, Iw2}.
For example, the $ \nu = 2/3 $ quantum Hall state (QHS) has two ground states with different spin polarizations \cite{6}.   
At the spin phase transition point of the $\nu =2/3$ QHS, an anomalous magnetoresistance $R_{xx}$ peak 
with hysteretic transport was observed by Kronm\"{u}ller et al. \cite{13}. 
The RF-irradiation experiment with a nuclear magnetic resonance frequency of the atoms 
in the quantum well has shown 
nuclear spin polarization to be the origin of the $R_{xx}$ peak \cite{4}. 
The relaxation measurement by Hashimoto et al. \cite{5} has suggested that the enhancement of the magnetoresistance at the $\nu =2/3$ QHS is proportional to the degree of nuclear spin polarization.
Thereafter, resistively detected nuclear magnetic resonance and relaxation measurements became excellent techniques 
for the study of the spin state of various exotic QHSs as such as the  Goldstone mode of the Skyrmion lattice near the $\nu =1$ QHS \cite{5}, 
the canted antiferromagnetic phase in the bilayer $\nu =2$ QHS \cite{36},
and the spin polarization in the even-denominator fractional $\nu =5/2$ non-Abelian QHS \cite{30}. 

Up to now, the mechanism of the $R_{xx}$ enhancement has been understood as follows: at the spin transition point, 
the two fractional quantum Hall states with different spin polarizations degenerate energetically in the composite Fermion model \cite{6,16}, 
and an electronic domain structure is formed \cite{13,8,3}. 
It is believed that when a current passes across a domain boundary, 
electron spins flip-flop scatter nuclear spins causing dynamic nuclear spin polarization (DNP), which in turn affects the domain formation and increases the length of the domain boundaries \cite{3,8}.
In addition, an anisotropy of the hysteretic transport  for the angle between current direction and in-plane magnetic field around the transition point in a tilted magnetic field was found \cite{Iwata}. This suggests that the
dynamics of electron spin domains is affected by the current direction for the in-plane field.   
However, details of the domain structure are still unclear, and 
the real cause of the magnetoresistance enhancement remains a riddle even now.

A previous experiment has shown that 
the resistance-enhanced state (RES) is not stable and decays over roughly several hundred seconds \cite{Iwata}.
In this work, the measurement of the resistance is performed much faster than the decay time of RES. 
In order to avoid the heating effect due to the current, we change the measuring current to a sufficiently small value. Moreover we measure the temperature dependence of the $R_{xx}$ of RES within a short time
after pumping the nuclear spin polarization with a large current of 60\,nA.  
With this new method, we find that the magnetoresistance of RES has a negative temperature gradient. 
By investigating the $R_{xx}$ and the Hall resistance $R_{xy}$ in the RES as a function of $\nu$, the RES is found to indicate a quantized Hall resistance. We explain the quantized Hall resistance as the property of the FQHS domain surrounded by the Anderson localized state.

The sample employed in our experiments consists of two 20\,nm-thick GaAs/AlGaAs 
quantum wells that have been processed 
into a 50\,$\mu$m-wide Hall bar with a voltage probe distance of 180 $\mu$m . 
We use the front layer and deplete electrons in the back layer.
This sample was previously used for the experiment on interlayer diffusion of the nuclear spins \cite{41a}. 
The electron density $n $ is controlled by the gate voltage. 
The low temperature mobility is $2.2\times 10^6$ cm$^2$/Vs with $n =2.0\times 10^{11}$\,cm$.^{-2}$
The sample is placed inside a mixing chamber filled with liquid helium 
in a dilution refrigerator 
with a base temperature of 62\,mK.
We measure the $R_{xx}$ and $R_{xy}$ with a standard low-frequency AC lock-in technique 
at a frequency of 37.7\,Hz.  
The magnetic field is applied by a superconducting magnet with a maximum field of 13.5\,T.
The measurement is performed at 7.19\,T; 
therefore, the phase transition of the $\nu=2/3$ QHSs just arise. 
A RuO$_2$ resistance calibrated in the magnetic field is used as a thermometer, and 
a heater is placed near the sample immersed 
in the liquid helium of the mixing chamber.

\begin{figure}[t]
\includegraphics[width=0.90\linewidth]{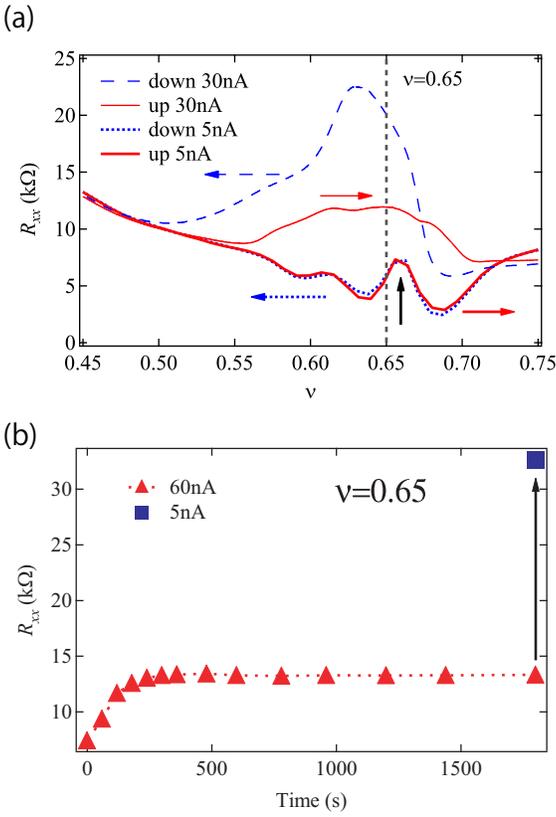}
\caption{ (Color online)
(a) Plot of the magnetoresistance versus filling factor 
around $\nu = 2/3$ at $n=1.8\times 10^{11}$ cm$^{-2}$.
The thick red (blue dotted) line represents a sweep from a lower (higher) to a higher (lower) filling factor at a current of 5\,nA.
The maximum in $R_{xx}$ (indicated by the up arrow) corresponds to the phase transition point of the spin polarized and unpolarized states.
The spin polarized and unpolarized $\nu =2/3$ QHSs are observed at the minimums of the left and right sides of the maximum, respectively.  
The solid red (broken blue) line represents a $\nu$  sweep in the increasing (decreasing) direction with 30\,nA current.
We find a clear hysteresis in $R_{xx}$.
(b) Time evolution of $R_{xx}$ at $\nu =0.65 $ point and $ I = $ 60\,nA.
After pumping, decreasing the current to 5\,nA yields two times higher $R_{xx}$. 
}
\end{figure}

Figure 1(a) shows the hysteretic behavior of $R_{xx}$ around the $\nu=2/3$ QHS.
The front gate voltage is swept to change $\nu$.
The thick red (blue dotted) line shows $R_{xx}$ as a function of $\nu$ around $\nu=2/3$ 
with an AC current of 5\,nA and a upward (downward) sweep speed of $d\nu /dt=1\times10^{-2} $\,s$^{-1}$.
No hysteretic behavior is seen, and the $R_{xx}$ peak of the spin phase transition point is clearly observed.
The thin red (blue broken) line represents the upward (downward) sweep with 
$d\nu /dt=1\times10^{-4} $\,s$^{-1}$ and a large current, $I=30$\,nA. 
The hysteretic behaviors of $R_{xx}$ are seen 
around the spin transition point with slow sweep and high current.

These results are similar to previous reports about the hysteretic behavior \cite{8,24,64}.
At $\nu=0.65$ (the dotted straight line in Fig. 1(a)) , 
where the enhancement of $R_{xx}$ is large, 
we perform DNP with a 60\,nA current continuously and at 1,800\,s change the current to 5\,nA.
By pumping nuclear spin polarization, $R_{xx}$ increases for about 500\,s and saturates. 
After pumping, sudden change of the measuring current to small value of 5 nA to avoid the self-heating effect, the $R_{xx}$ is enhanced  
by a factor of two (the blue square in Fig.1(b)) 

\begin{figure}[t]
\includegraphics[width=1\linewidth]{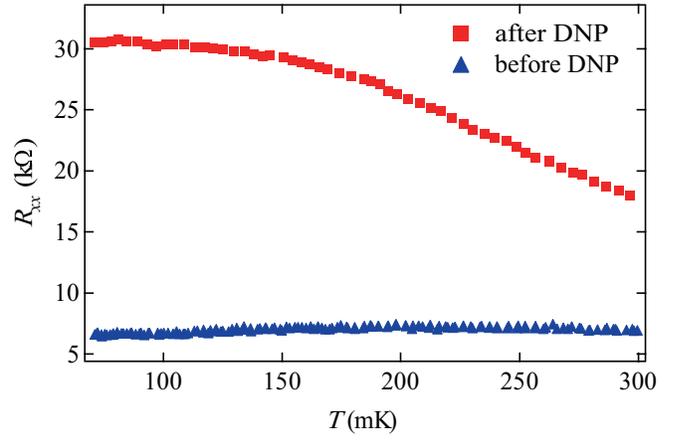}
\caption{ (Color online)
Temperature dependence of $R_{xx}$. 
Blue triangles are the QHS data before DNP, which indicates an activation energy gap of $\Delta=15$\,mK.
The red squares are the data after DNP with a current of 60\,nA and 37.7\,Hz AC current. 
The notable feature is that the data after DNP indicate a negative gradient for the temperature dependence.
}
\label{f2}
\end{figure}

In the next step, to investigate the temperature dependence of the $R_{xx}$ in the RES, 
we increase quickly the temperature of the mixing chamber 
containing the sample with the heater within 50\,s after DNP. 
The solid blue triangles represent the temperature dependence of $R_{xx}$ before DNP.
The slope gives an activation energy gap of about $\Delta=15$\,mK
when analyzed with the Arrhenius formula.
The red squares are the data after DNP with a current of 60 nA.
Surprisingly, the derivative of $R_{xx}$ with respect to the temperature is negative. In a previous experiments \cite{13}, the current dependence of $R_{xx}$ for $\nu =2/3$ was investigated, but the insulating behavior was not observed, because the strong DNP was induced by the high current, 
considered to be simultaneously extinguished the $R_{xx}$ decrement owing to the heating of the high-current measurements.
There are two possible explanations for this temperature dependence: one is trivia, in 
that the nuclear spin polarization rapidly relaxes owing to the temperature increase,
and the other explanation is the intrinsic thermal property of the RES,
implying the transition of the QHS into an insulating phase.

To determine which explanation is more appropriate, we measure 
the temperature dependence of $R_{xx}$ in a different way, using the self-heating effect of the probing current.   
Figure 3(a) shows the $R_{xx}$ in the RES as a function of the probing current.
The reference current for measuring $R_{xx}$ is 5\,nA. The measurement procedure is as follows:
1) set $\nu =0.65$ and pump DNP at 60\,nA for 1,800\,s;
2) decrease the current to 5\,nA and measure $R_{xx}$;
3) change the current to arbitrary values $ I_{\text{meas}} $;
4) measure $R_{xx}$ for several different values of $ I_{\text{meas}} $ 
in a total time of about 20\,s;
5) decrease the current to 5\,nA and measure $R_{xx}$;
6) reset the RES using the Goldstone mode of the Skyrmion crystal state, where the polarization quickly relaxes to the equilibrium value\cite{5} when the filling factor is changed to $\nu = $ 0.85  ;
7) go back to 1) and  repeat.
Because the values measured at the reference current in steps 2) and 5) are in good agreement, 
the RES does not change during the measurements for the various currents.    
The sample temperature increases by a extremely small electric power because the heat capacity of the 2DES is very small.\cite{93} 
Moreover, the temperature of the surrounding liquid helium is maintained at a low temperature 
because the Kapitza resistance between the semiconductor and the liquid helium is very high.\cite{95}
Thus, in this situation, the assumption that the probing current changes only the temperature of the sample is plausible.

\begin{figure}[t]
  \includegraphics[width=1.0\linewidth]{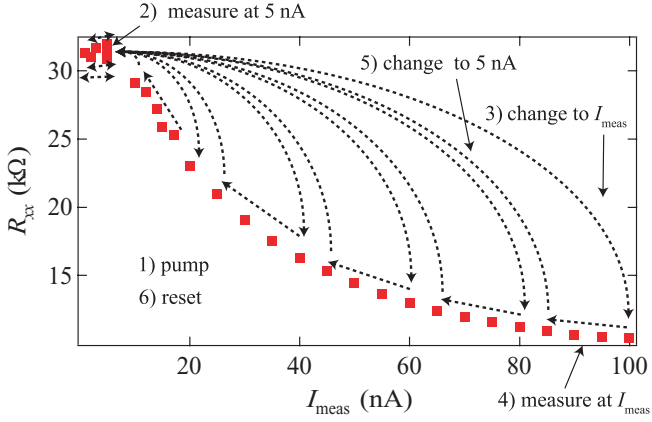}
  \caption{(Color online)
Temperature dependence of $R_{xx}$ using the self-heating effect of the probing current. 
The arrows indicate the measuring sequences. We refer the resistance to 5\,nA before and after the measurement of the various currents to verify the persistence of the RES.
}
  \label{f3}
\end{figure}

The temperature of the 2DES is estimated under the assumption that the same energy input 
yields the same temperature increase in any kind of QHS.
To get the relation between the power $ P $ and the sample temperature $ T $, we use the $\nu =1/3$ QHS.
The procedure is as follows: 
1) we measure the temperature dependence of $R_{xx}$ 
at 1\,nA by increasing the temperature of the mixing chamber;
2) we measure $R_{xx}$ for various currents;
and 3) we compare 1) and 2) to determine the power dependence of the sample temperature $T_\text{eff}(P)$.
The inset of Fig. 4(a) shows the plot of the $T_\text{eff}$ against the resistive heating $P=R_{xx}I^2$. 
Fig. 4(b) shows the result of calibrating the electric power to the sample temperature for Fig. 3. 
We assume that same  $R_{xx}$ value is associated with the same temperature value, for instance, 300\,mK in Fig. 2 corresponds to 240\,mK in Fig. 4(b). Here the temperature difference is 20\,\%.
One cause of this difference is likely to be the lack of the thermal equilibrium between the sample and the thermometer, 
which is a potential difficulty, in any experiment using a heater. 
As for other cause, the calibration curve of $T_\text{eff}(P)$ may be inaccurate at the insulating phase. 
The inspections of the calibration curve are nice result checked in the $R_{xx}$ of any quantum Hall state (Fig. 4(a)).
But it may be necessary to inspect whether it is possible under the conditions to have the characteristic of an insulating state, because the RES indicate the negative temperature gradient.
Although there is a discrepancy of 20\,\%, both measurements indicate that the temperature
derivative of the magnetoresistance is negative.
 
\begin{figure}
\begin{center}
  \includegraphics[width=0.8\linewidth]{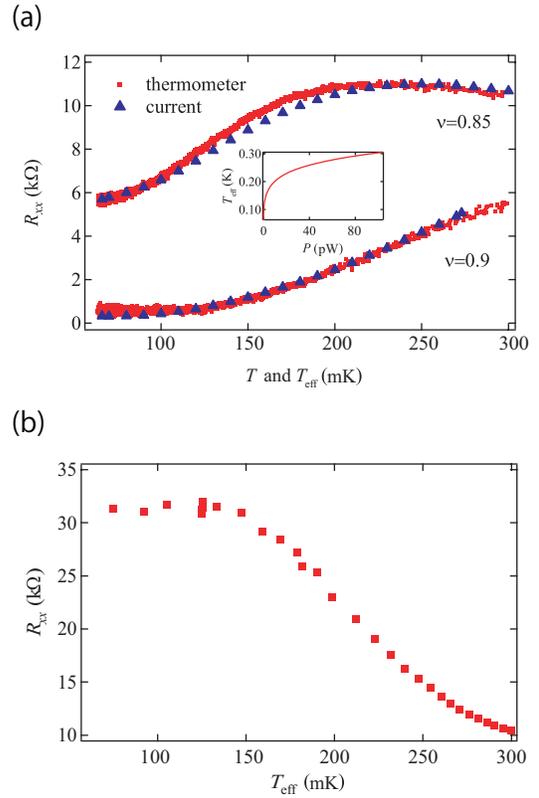}
\end{center}
  \caption{
(Color online)
(a) Result of calibrating electric power to the converted temperature. The inset shows the power dependence of the sample temperature $T_\text{eff}(P)$, by constructing the relation between heating power and temperature using the $R_{xx}$ of the $\nu =1/3$ QHS. The main figure compares two different measuring methods by testing at any filling factor ($\nu=0.85, 0.9$). Red squares are measured by an external thermometer and heater. Blue triangles are measured by self-heating of probing current and converting from power to temperature. 
(b) Conversion of the horizontal axis of Fig. 3 from current to temperature.
The  current is converted using by $T_\text{eff}(P)$ function in inset of this figure (a). 
}
\label{f4}
\end{figure}

To clarify the intrinsic properties of  the RES at $\nu =2/3$, we measure the magnetoresistance and the Hall resistance simultaneously in the RES 
as a function of the filling factor.
In Fig. 5, the $R_{xx}$ before and after DNP are plotted for measuring current values of 5\,nA and 60\,nA.
The broken vertical line is located at $ \nu=0.65$, where the DNP is carried out.
For the entire range of $ \nu$, the $R_{xx}$ before DNP with 5\,nA is lower than the one with 60\,nA. Since the sample temperature is proportional to the heating effect by the DNP current, the QHS is developed for the entire range of $\nu$ before DNP.
However, after DNP, the differential coefficient of $R_{xx}$ to the temperature takes negative values for $ 0.5 \leq \nu \leq 0.68 $; that is, the novel insulator state arises.
In Fig. 5(b), the Hall resistance coincides approximately with a quantized value of $3h/2e^2$,
not only in the QHS but also in the insulating state of the RES.

\begin{figure}
\begin{center}
  \includegraphics[width=1.0\linewidth]{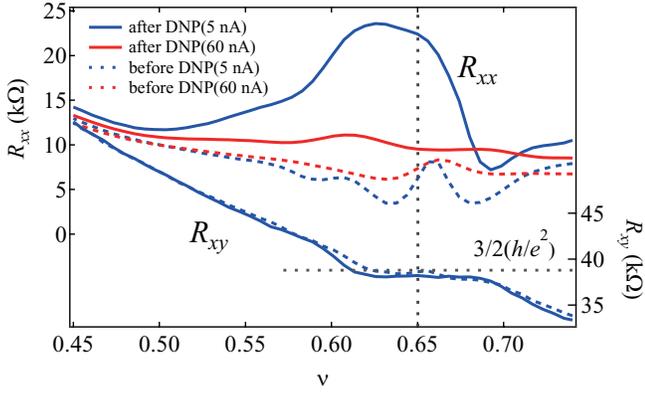}
\end{center}
  \caption{
(Color online)
$R_{xx}$ before and after DNP are illustrated for 5\,nA and 60\,nA measuring currents as a function of the filling factor.
$R_{xy}$ before and after DNP are illustrated for 5\,nA measuring current as a function of the filling factor.
The broken vertical lines indicate $ \nu $ for DNP. 
After DNP, the insulator state arises for a very wide $ \nu $ range.
The Hall resistance is quantized at a value of $3h/2e^2$ even in the insulating state of the RES. 
}
\label{f4}
\end{figure}

\begin{figure}
\begin{center}
\includegraphics[height=.23\textheight]{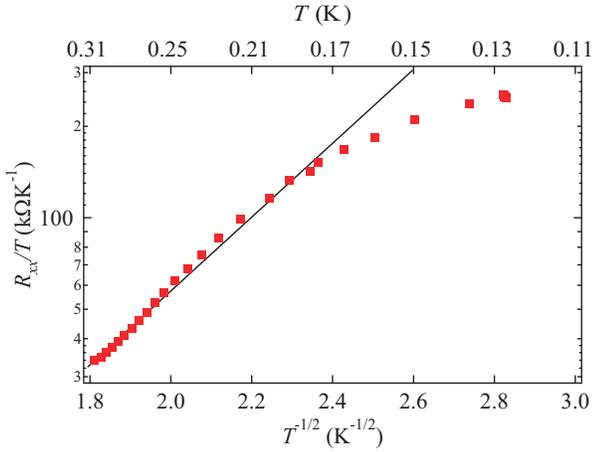}
\end{center}
\caption{
(Color online)
Logarithm of $R_{xx}/T$ versus $1/T^{-1/2}$. 
Data are the same as in Figs. 3. At a low temperature, 
data deviate from dependence on $T^{-1/2}$. 
}
\end{figure}

Here, we discuss the mechanism for the insulating properties of the RES. 
One of the candidates for the insulating phase is a Wigner crystal 
that manifests itself at $\nu <1/7$ using the high mobility samples.
\cite{47, 49, 46, 98}
The RES is induced by effectively increasing the disorder caused by the nuclear spin polarization
that acts as electron scattering centers. Since this disorder
destroy the long-range order of the Wigner crystal, the
possibility of the Wigner crystal forming in this experiment is
quite low. 
Another candidate is a stripe phase, which has the
same temperature dependence as the insulating phase.
\cite{72, 52}
However, it is unlikely that the stripe phase will appear,
because it grows in a unidirectional potential or an
in-plane magnetic field only in high mobility samples.
The Mott insulator is also a candidate for the insulating phase.
Since $R_{xy}$ diverges in the Mott insulator, this state is eliminated as a candidate\cite{99}.
Now we try to find another possibility. 
Kagalovsky and Vagner \cite{34} investigated localization-delocalization transition in quantum Hall systems with a random field of nuclear spins acting on 2DES via hyperfine contact interaction.
Because DNP affects the additional confining potential of the electrons , 
we attempt to explain the generating mechanism of the insulator by the Anderson localization.
In the Anderson insulator, the electron conducts by variable-range hopping,
and follows the form for a quantum Hall regime:\cite{89, 90}
\begin{equation}
\frac{R}{R_0(T)}=\exp \left(\frac{T_0}{T}\right)^{1/2} \mspace{20mu} 
\end{equation}
with
\begin{equation}
k_BT_0=C\frac{e^2}{4\pi\epsilon\epsilon_0\xi} \mspace{20mu}  
\end{equation}
and
$R_0(T)=T/S_0$, where $\xi$ is the
 localization length, $S_0$ the constant conductivity prefactor, 
and $C$ the hopping constant.
We  obtain the characteristic temperature $T_0=7.0$\,K
by replotting the data of Figs. 4 as $\log R/T$ versus $T^{-1/2}$ (Fig. 6).
The localization length can be derived using Eq. (2),
where $C=6$ in a 2DES\cite{96}.
Thus the localization length is estimated to be approximately $1.1$\,$\mu$m.
As the localization length is much smaller than the size of the Hall bar, 
a strongly localized system dominates, namely an insulating state.      

Lastly, we discuss the quantization of the Hall resistance in the insulating state. 
Shimshoni and Auerbach\,\cite{Shim} interpreted the coexistence of the quantized $R_{xy}$ and insulating $R_{xx}$ with a model in which FQHS ``puddles" are surrounded by an insulating phase 
and connected to one another by tunnel junctions. 
In the $\nu = 2/3$ system, FQHS puddles connected through the insulating phase by the tunneling effect. 
As seen in Fig. 5, the $R_{xy}$ quantization is not stable over the entire $\nu$ range.
For large $\nu $ ($0.62<\nu<0.69$), where $R_{xy}$ is quantized, the FQHS puddle is formed in the Anderson insulator phase, and for low $\nu$ ($\nu<0.62$), $R_{xy}$ deviates from the quantized value; that is, the FQHS puddles disappear, and the Anderson insulating phase extends all over the 2DES. 
A similar state has been observed in the low mobility 2DES \cite{Hilke} and is called the quantized Hall insulator.

In summary, we measured the temperature dependence of the enhanced magnetoresistance
pumped by the electric current.
Our results indicate that the enhanced resistance state is 
a novel insulating phase that appears as result of a phase transition from the QHS induced by dynamic nuclear spin polarization.
Anderson localization is enhanced by a number of tiny polarized nuclear spins acting as impurities.
By investigating the magnetoresistance and the Hall resistance in the RES, the insulating state is found to be the quantized Hall resistance. We explained the quantized Hall resistance by a model in which FQHS puddle are surrounded by an Anderson insulating phase.
 
\begin{acknowledgments}
We are thankful to N. Kumada, K. Muraki,  
and Y. Hirayama  
for not only providing us with a high mobility sample but also for many fruitful discussions. 
We are also grateful to Y. Ono, T. Nakayama, S. Takagi, Z. F. Ezawa, N. Shibata, T. Arai, and K. Iwata for useful discussions. 
This research was supported in part by MEXT/JSPS KAKENHI Grant Number 24540319, 24540331, 25103722
and the Grant-in-Aid for the Global COE Program
"The Next Generation of Physics, Spun from Universality and Emergence."
\end{acknowledgments}

\end{document}